\documentstyle[sprocl,cite]{article}

\def\Journal#1#2#3#4{{#1} {\bf #2}, #3 (#4)}

\def\NPB{{\em Nucl.Phys.}B}
\def\PLB{{\em Phys.Lett.}B}
\def\PRL{\em Phys.Rev.Lett.}
\def\PRD{{\em Phys.Rev.}D}

\def\JETP{\em JETP}

\def\PAN{\em Phys.At.Nucl.}

\newcommand{\eq}[1]{\begin{equation}#1\end{equation}}

\begin{document}

\title{Parametric Resonance of Neutrino Oscillations in \ \ \ \ \ \
Electromagnetic Wave}
\author{ M. DVORNIKOV, A.STUDENIKIN\footnote{\normalsize E-mail: studenik@srdlan.npi.msu.su}}

\address{Department of
Theoretical Physics, Moscow State University,
119899 Moscow, Russia}


\maketitle



Experimental studies of
solar, atmospheric and reactor neutrinos over the past several years
provide almost certain indications that neutrinos oscillate, have masses
and mix. These new properties of neutrinos, if confirmed by better
statistics in the proceeding and forthcomming
experiments, will require a significant departure from
the Standard Model. It is well known [1] that massive neutrinos can have
nonvanishing magnetic moment. For example, in the Standard Model supplied with
$SU(2)$-singlet right-handed neutrino the one-loop radiative correction
generates neutrino magnetic moment which is proportional to neutrino mass.
There are plenty of models [2] which predict much large magnetic moment for
neutrinos.

In the light of new developments in neutrino physics, to get better
understending of the electromagnetic properties of neutrinos
is an important task. In general, there are two aspects of this problem.
The first one is connected with values of neutrino electromagnetic form
factors.
The second aspect of the problem implies consideration of influence of
external electromagnetic fields, which could be presented in various
invironments, on neutrino possessing nonvanishing electromagnetic moments.
The most important of the latter are the magnetic and electric dipole moments.

The majority of the previously performed studies of neutrino conversions
and oscillations in electromagnetic fields deal with thecase of
constant transversal magnetic field or constant transversal twisting
magnetic fields (see references [1-18] of paper [3]). Recently we have
developed [3,4] the Loretz invariant formalizm for description of
neutrino spin evolution that enables one to consider neutrino oscillations
$ $ in the presence of an arbitrary electromagnetic fields. Within the
proposed approach it becomes possible to study neutrino spin evolution in
an electromagnetic wave and the new types of resonances in the neutrino
oscillations in the wave field and in some other combinations of fields have
been predicted.

The aim of this paper is to continue the study of the neutrinio spin
oscillations in the circular polarized electromagnetic wave. Here we consider
the case when the wave amplitude is not constant but is periodically vareing
function of time and show for the first time that under certain conditions
the parametric resonance of neutrino oscillations can occur in such a system.
Note that in order to investigate this fenomenon we have to use the Lorentz
invariant formalizms for neutrino evolution that have been developed in [3,4].
The possibility of the paramentric resonance of neutrino flavour oscillations
in matter with periodic variation of density was considered previously in [5].
It should be also pointed here that the conditions for a total neutrino
flavour conversion in a medium consisting of two or three constant density
layers were derived in [6]. For the recent discussion on the physical
interpretations of these two mechanizms of increasing of neutrino conversion
see
refs. [7] and [8].

Let us consider evolution of a system $\nu =(\nu_{+},\nu_{-})$ composed of
two neutrinos of different helicities in presence of a field of circular
polarized electromagnetic wave with periodically vareing amplitude. Here
neutrinos $\nu_{+}$ and $\nu_{-}$ correspond to positive and negative helicity
states that are determined by projection operators
\eq{P_{\pm}={{1}\over{2}}
\left(1\mp {{({\vec \sigma}{\vec p})}\over{|\vec p|}}
\right).}
In general case it is important to distinguish helicity states
$\nu=(\nu_{+},\nu_{-})$ and chirality states $\nu=(\nu_{L},\nu_{R})$.
The latter are determined by the projection operators $P_{L,R}=(1\mp
\gamma _{5})/2$.
The evolution of such a system is given [3,4] by the following
Schr$\ddot o$dinger type equation

\eq{i{{\partial \nu}\over{\partial t}}=H\nu, \ \ \
H={\tilde \rho}\sigma_{3}+E(t)(\sigma_1 \cos \psi -
\sigma_2 \sin \psi ),}
$${\tilde \rho}=-{{V_{eff}}\over{2}}+{\Delta{m}^{2}A \over 4E}, \ \ \
E(t)=\mu B(t)(1-\beta\cos \phi).$$
Here the three parameters, $A=A(\theta)$ being a function of vacuum mixing
angle, $V=V(n_{eff})$ being the difference of neutrino effective
potentials in matter, and $\Delta m^{2}$ being the neutrino masses
squired difference depend on the considered type of neutrino conversion
process. The electromagnetic field is determined in the laboratory frame of
reference by its frequency, $\omega$, the fase at the point of
the neutrino location,
$\psi=g\omega t(1-\beta/\beta_{0}\cos \phi)$,($g={\pm}1$), and the
amplitude, $B(t)$, which is a function of time. The wave speed in matter could be
less than the speed of light in vacuun $(\beta_0 \leq 1$), and $\phi$ is
the angle between the neutrino speed $\vec \beta$ and the direction of
the wave propagation. In the derivation of the Hamiltonian of eq.(2) terms
proportional to
${{1}\over{{\gamma}^{2}}}\ll{1}$,
$\gamma=(1-\beta^{2})^{-1/2}$,
and also an oscillating function of time in the diagonal part are neglected
(for details see [4]).

In order to stady fenomenon of the parametric resonance of neutrino spin
oscillations in such a wave we suppose that the amplitude $B(t)$ is given by
\eq{B(t)=B(1+hf(t)), \ \ \ (|h|\ll 1),}
where $f(t)$ is an arbitrary function of time and $h$ is a small
dimensionless quantity of not fixed sign. It is convenient to introduce the
evolution operator which determines the neutrino state at time $t$
$$\nu(t)=U(t)\nu(0)$$
if the initial neutrino state was $\nu(0)$. Using the Hamoltonian of eq.(2)
we getthe following equation for the evolution operator:
\eq{{\dot U}(t)=i[-{\tilde \rho}{\sigma_{3}}+(-E_{0}+\varepsilon f(t))
(\sigma_1 \cos \psi - \sigma_2 \sin \psi )]U(t),}
where c$\varepsilon=-E_{0}h$. In analogy with the cesa of the electromagnetic
wave with nonvareing amplitude [3,4] the solurion of eq.(4) can be written
in the form
\eq{U(t)=U_{\vec e_{3}}(t)U_{\vec l}(t)F(t),}
where
$U_{\vec e_{3}}(t)=\exp(i\sigma_{3}{{{\dot \psi}t}\over{2}})$
is the rotation operator around the axis
${\vec e}_{3}$ which is parallel with the direction of the neutrino
propagation, and
$U_{\vec l}(t)=\exp(i{\vec \sigma}{\vec l}t)$
is the rotation operator around the vector
${\vec l}=(-E_{0},0,{\tilde \rho}-{{\dot \psi}\over{2}})$.
It should be noted that the solution of eq.(4) for the case of constant
amplitude of the wave field $(\varepsilon=0)$ is given by the operator
$U_{0}(t)=U_{\vec e_{3}}(t)U_{\vec l}(t)$.

From (4) and (5) it follows that the equation for the operator $F(t)$ is
\eq{{\dot F}(t)=i\varepsilon
H_{\varepsilon}(t)F(t),}
where
$$H_{\varepsilon}(t)=({\vec \sigma}{\vec y}(t))f(t),$$

$${\vec y}=\big(1-2\lambda_{3}^{2}\sin^{2} \Omega t,
\ \lambda_{3}\sin 2\Omega t,\
2\lambda_{1}\lambda_{3}\sin^{2} \Omega t\big),$$
and the unit vector $\lambda$ is given by its components in the unit orthogonal
basis ${\vec \lambda}=
\lambda_{1}{\vec e}_{1}+\lambda_{2}{\vec e}_{2}+\lambda_{3}{\vec e}_{3}=
{{\vec l}\over{\Omega}}, \Omega=|{\vec l}|$. The detailed analysis of
evalution of the solutionof eq.(6) can be found in [9]. Here we comment only
on the main steps. Using the smallness of $\varepsilon$ we expand the solution
of eq.(6) in powers of this parameter:
\eq{F=\sum_{k=0}^{\infty}\varepsilon^{k}F^{(k)},}
where
$F^{(0)}={\hat 1}$ is a unit matrix. For operators
$F^{(k)}$ the recurrent formula is strightforward:
\eq{F^{(k+1)}(t)=
i\int\limits_{0}^{t}H_{\varepsilon}(\tau)F^{(k)}(\tau)d\tau.}
Skipping further technical details to the first order in $\varepsilon$ we get
\eq{F(t)={\hat 1}+i\varepsilon({\vec \sigma}{\vec x}(t))+
O(\varepsilon^{2}),}
where
$${\vec x}(t)=
\int\limits\limits_{0}^{t} {\vec y}(\tau)f(\tau)d\tau.$$
Thus for the probability of neutrino conversion
$\nu_{i-}\leftrightarrow\nu_{j+}$
we get
\eq{
P_{ij}=
\big|<\nu_{+}|U_{\vec e_{3}}(t)U_{\vec l}(t)F(t)|\nu_{-}>\big|^{2}=}
$$\lambda_{1}^{2}\sin^{2}\Omega t+2\varepsilon \lambda_{1}
\big(x_{1}(t)\cos \Omega t+\lambda_{3}x_{2}(t)\sin \Omega t\big)
\sin \Omega t.$$
Note that $\Omega$ is nothing but the mean osillation frequency of the
neutrino system.

For further evaluation of solution of eq.(6) we have to specify the
form of the function $f(t)$. As it is mentionaed above, the purpose
of this study is to examine the case when the parametric resonance for neutrino
oscillations in electromagnetic wave with vareing amplitude could appear.
Having in mind a simple analogy ( see, for example, [7]) between oscillations
in neutrino system and oscillations of a classical pendulum [10], it is
resonable to suppose that the principal parametric resonance apperas when the
amplitude modulation function $f(t)$ is oscillating in time with frequency
approximately equal to the twice mean oscillation frequency of the system.
That is why we choose the function $f(i)$ to be
\eq{f(t)=\sin 2{\Omega}t}
and get for the neutrino oscillations probability
\eq{P_{ij}=\big[\lambda_{1}^{2}+\varepsilon
\lambda_{1}\lambda_{3}^{2}t+ {{\varepsilon \lambda_{1}}\over{\Omega}}
\big(1-{{\lambda_{3}^{2}}\over{2}}\big)\sin 2\Omega t\big]
\sin^{2}\Omega t.}
It follows that in the case $\lambda_{1}\varepsilon>0$ the second term
increases with increase of time $t$, so that the amplitude of the
neutrino conversion probability may becomes close to unity. This is the effect
of the parametric resonance in neutrino system in the elctromagnetic wave
with modulated amplitude that may enhance neutrino oscillation amplitude even
for rather small values of the neutrino magnetic moment $\mu$ and strengh
of the electromagnetic field and also when the system is
far away from the region of ordinary spin (or spin-flavour) neutrino resonance.

In conclusion, let us consider the case when parameters of the neutrino system
are far beyond the region of the ordinary resonance. Then the following
condition is valid: $\lambda_1 \ll \lambda_3$, and the maximal neutrino
conversion probability is small for the case of nonvareing $(h=0)$ amplitude
of the electromagnetic field
\eq{P_{ijmax}(h=0)={{l_{1}^{2}}\over{l_{1}^{2}+l_{3}^{2}}}\ll 1.}
If the amplitude of the field is modulated in accordance with eq.(11)
the estimation for the critical time $t_{cr}$ for which the probability
$P_{ij}$
could become close to unity for an arbitrary values of the neutrino mixing
angle, masses, energy, and density of matter gives
\eq{t_{cr}\approx{{1}\over{\varepsilon n_{1}}}.}
It means that the parametric resonance enhancenment of the neutrino
oscillations occurs after neutrinos travel a distance
\eq{L={{\Omega}\over{|h|[\mu B(1-\beta\cos \phi)]^{2}}}.}

One of us (A.S.) should like to thank Jorge Dias de Deus, Ana Mourao,
Paulo Sa and all the organizers of the meeting for hospitality


\section*{References}
\begin{thebibliography}{99}

\bibitem{1} B.Lee, R.Shrock, \Journal{\PRD}{16}{1444}{1977};
K.Fujikawa, R.Shrock, \Journal{\PRL}{45}{963}{1980}.

\bibitem{2} J.Kim, \Journal{\PRD}{14}{3000}{1976};
            M.Beg, W.Marciano, \Journal{\PRD}{17}{1395}{1978};
            M.Voloshin, M.Vysotsky, L.Okun,
            \Journal{\JETP}{64}{446}{1986};
            M.Fukugita, T.Yanagida, \Journal{\PRL}{58}{1807}{1987}.

\bibitem{3} A.Egorov, A.Lobanov, A.Studenikin, \Journal{\PLB}{491}{137}{2000};
           hep-ph/9910476.

\bibitem{4} A.Egorov, A.Lobanov, A.Studenikin, in : A.Mourao, M.Pimento, P.Sa
            (Eds.), New Worlds in Astroparticle Physics, World Scientific,
            Singapore, 1999, p.153; hep-ph/9902447.

\bibitem{5} V.Ermilova, V.Tsarev, A.Chechin, Short Notices of the Lebedev
            Institute, v. 5, p. 26, 1986; E.Akhmedov,
            \Journal{SJNP}{47}{301}{1988}; P.Krastev, A.Smirnov,
            \Journal{PLB}{226}{341}{1989}.

\bibitem{6} M.Chizov, S.Petcov, \Journal{\PRL}{83}{1096}{1999};
            hep-ph/9903424, \Journal{\PRD}{63}{073003}{2001};
            (E) \Journal{\PLB}{444}{584}{1998};
            S.Petcov \Journal{\PLB}{434}{321}{1998}.

\bibitem{7} E.Akhmedov, \Journal{\NPB}{538}{25}{1999}; E.Akhmedov, A.Smirnov,
            hep-ph/9910433.

\bibitem{8} M.Chizov, S.Petcov, hep-ph/0003110.


\bibitem{9} M.Dvornikov, A.Studenikin, \Journal{\PAN}{64}{to be published}{2001}.

\bibitem{10} L.Landau, E.Lifshits, Course of Theoretical Physics, V.1,
Mechanics, 3rd ed., Pergamon Press, London, 1976.



\end{thebibliography}
\end{document}